\documentclass[useAMS,usenatbib]{mnras}
\pdfoutput=1

\usepackage[pdftex]{graphicx,color}
\usepackage[T1]{fontenc}

\definecolor{urlblue}{rgb}{0,0,0.9}
\definecolor{linkblue}{rgb}{0,0,.8}
\definecolor{linkgreen}{rgb}{0,0.45,0}
\definecolor{linkpurple}{rgb}{0.7,0.0,0.4}
\definecolor{linkorange}{rgb}{0.7,0.1,0.0}

\usepackage{amsmath, amssymb}
\usepackage{cuted}
\usepackage[utf8]{inputenc}
\usepackage{lmodern}
\usepackage[english]{babel}
\usepackage{enumerate}
\usepackage[normalem]{ulem}
\usepackage{enumitem}

\setlist[enumerate]{wide=0pt, widest=99,leftmargin=\parindent, labelsep=* } 

\definecolor{vale}{rgb}{0,0.5, 1.}

\graphicspath{{./figs/}}

\AtBeginDocument{\hypersetup{
linkcolor=linkblue,
citecolor=linkorange,
urlcolor=urlblue}}

\providecommand{\eprint}[1]{\href{http://arxiv.org/abs/#1}{#1}}
\providecommand{\adsurl}[1]{\href{#1}{ADS}}

\usepackage{ifthen}\def\eprinttmp@#1arXiv:#2 [#3]#4@{\ifthenelse{\equal{#3}{x}}{\href{http://arxiv.org/abs/#1}{#1}}{\href{http://arxiv.org/abs/#2}{arXiv:#2} [#3]}}
\renewcommand{\eprint}[1]{\eprinttmp@#1arXiv: [x]@}

\title[Model-independent radial BAO]{A first model-independent radial BAO constraint from the final BOSS sample}

\author[Marra and Chirinos Isidro]{
Valerio Marra$^{1}$
and Eddy G.~Chirinos Isidro$^{2}$
\\
$^{1}$Núcleo Cosmo-ufes \& Departamento de Física, Universidade Federal do Espírito Santo, 29075-910, Vitória, ES, Brazil\\
$^{2}$PPGFis, Universidade Federal do Espírito Santo, 29075-910, Vitória, ES, Brazil
}

\date{Accepted XXX. Received YYY; in original form ZZZ}

\pubyear{2017}

\begin{document}
\label{firstpage}
\pagerange{\pageref{firstpage}--\pageref{lastpage}}

\maketitle

\begin{abstract}
Using almost one million galaxies from the final Data Release 12 of the SDSS's Baryon Oscillation Spectroscopic Survey, we have obtained, albeit with low significance, a first model-independent determination of the radial BAO peak with 9\% error: $\Delta z_{\rm BAO}(z_{\rm eff}=0.51)= 0.0456 \pm 0.0042$.
In order to obtain this measurement, the radial correlation function was computed in  7,700 angular pixels, from which mean correlation function and covariance matrix were obtained, making the analysis completely model independent.
This novel method of obtaining the covariance matrix was validated via the comparison with 500 BOSS mock catalogs.
This $\Delta z_{\rm BAO}$ determination can be  used to constrain the background expansion of exotic models for which the assumptions adopted in the standard analysis cannot be satisfied.
Future galaxy catalogs from J-PAS, DESI and Euclid are expected to significantly increase the quality and significance of model-independent determinations of the BAO peak, possibly determined at various redshift and angular positions.
We stress that it is imperative to test the standard paradigm in a model-independent way in order to test its foundations, maximize the extraction of information from the data, and look for clues regarding the  poorly understood dark energy and dark matter.
\end{abstract}


\section{Introduction}\label{intro}

Thanks to the increasing wealth of astronomical data it started to be possible to analyze the cosmos with model-independent analyses.
The basic idea is to extract information from data without assuming the properties of the energy content of the universe nor, if possible, the overall spacetime structure \citep{Stebbins:2012vw}.
While model-independent analyses may be less constraining as compared to standard analyses, they have the advantage that they can be used to verify basic assumptions such as homogeneity and isotropy and to analyze exotic models for which standard results could not be used.
Indeed, it is becoming increasingly difficult to analyze alternative  models as the standard model of cosmology is more and more at the heart of the ever more complex data reduction pipelines.

Examples of cosmology-independent measurements are Supernova Ia model-independent calibrations \citep{Hauret:2018lnj}, expansion-rate determinations via redshift drift \citep{Martins:2016bbi} or the differential age evolution of cosmic chronometer \citep{Moresco:2018xdr} and their model-independent reconstructions \citep[see, e.g.,][and references therein]{Marra:2017pst}.
A model-independent ``low-redshift standard ruler'' (which, within the standard model, coincides with the sound horizon at radiation drag) was obtained by \cite{Verde:2016ccp}, model-independent CMB constraints were obtained by \cite{Vonlanthen:2010cd}, and ``standard sirens'' -- gravitational wave detections with electromagnetic counterparts -- can provide cosmology-independent determinations of the luminosity-distance-redshift relation \citep{TheLIGOScientific:2017qsa}.

Here, we will focus on constraints from Baryon Acoustic Oscillations (BAO).
The standard  analysis \citep[see][]{Alam:2016hwk} provides BAO measurements which are ``pseudo model-independent'' as they can easily be scaled to trial cosmologies using the Alcock-Paczynski distortion.
For this method to be valid the trial cosmology must be ``sufficiently close'' to 
the fiducial model.
For example, the 2-point correlation function template and the reconstruction technique may not be valid for, e.g., non-standard dark energy models.
In order to overcome these limitations \cite{Anselmi:2015dha} introduced the ``linear point'', a model-independent BAO standard ruler which can be used also for cosmologies different from $\Lambda$CDM \citep[see also][]{Anselmi:2018hdn,Anselmi:2017zss}.
Here, we wish to use a completely model-independent method, whose results could be used to constrain even the most exotic models such as inhomogeneous metrics.
As we will see, the price will be a much lower significance and accuracy with respect to the standard and linear-point analyses.

Model-independent angular BAO constraints were discussed, for example, by \cite{Sanchez:2010zg} in the context of a DES-like photometric redshift survey, and obtained using galaxy \citep{Alcaniz:2016ryy,Carvalho:2017tuu} and quasar \citep{deCarvalho:2017xye} catalogs of the the Baryon Oscillation Spectroscopic Survey (BOSS) of the Sloan Digital Sky Survey (SDSS).
\cite{Sanchez:2012eh} proposed a simple recipe in order to  measure the radial BAO in a model independent way. It was applied to a large Euclid-like $N$-body simulation and it was shown to be able to correctly obtain the BAO scale relative to the cosmology of the simulation.
Here, we extend the methodology introduced in \cite{Sanchez:2012eh} in order to apply it to the final Data Release 12 of the BOSS survey \citep{Eisenstein:2011sa}.

This paper is organized as follows. After describing the data in Section~\ref{data}, we describe  the method in Section~\ref{method}, placing particular emphasis on the use of galaxy and angular pixel weights and on the estimation of the covariance matrix. In Section~\ref{results} we show our results---the determination of the BAO feature in redshift space.
We conclude in Section~\ref{conclusions}.

\section{Data}\label{data}

\begin{figure}
\centering
\includegraphics[width= \columnwidth]{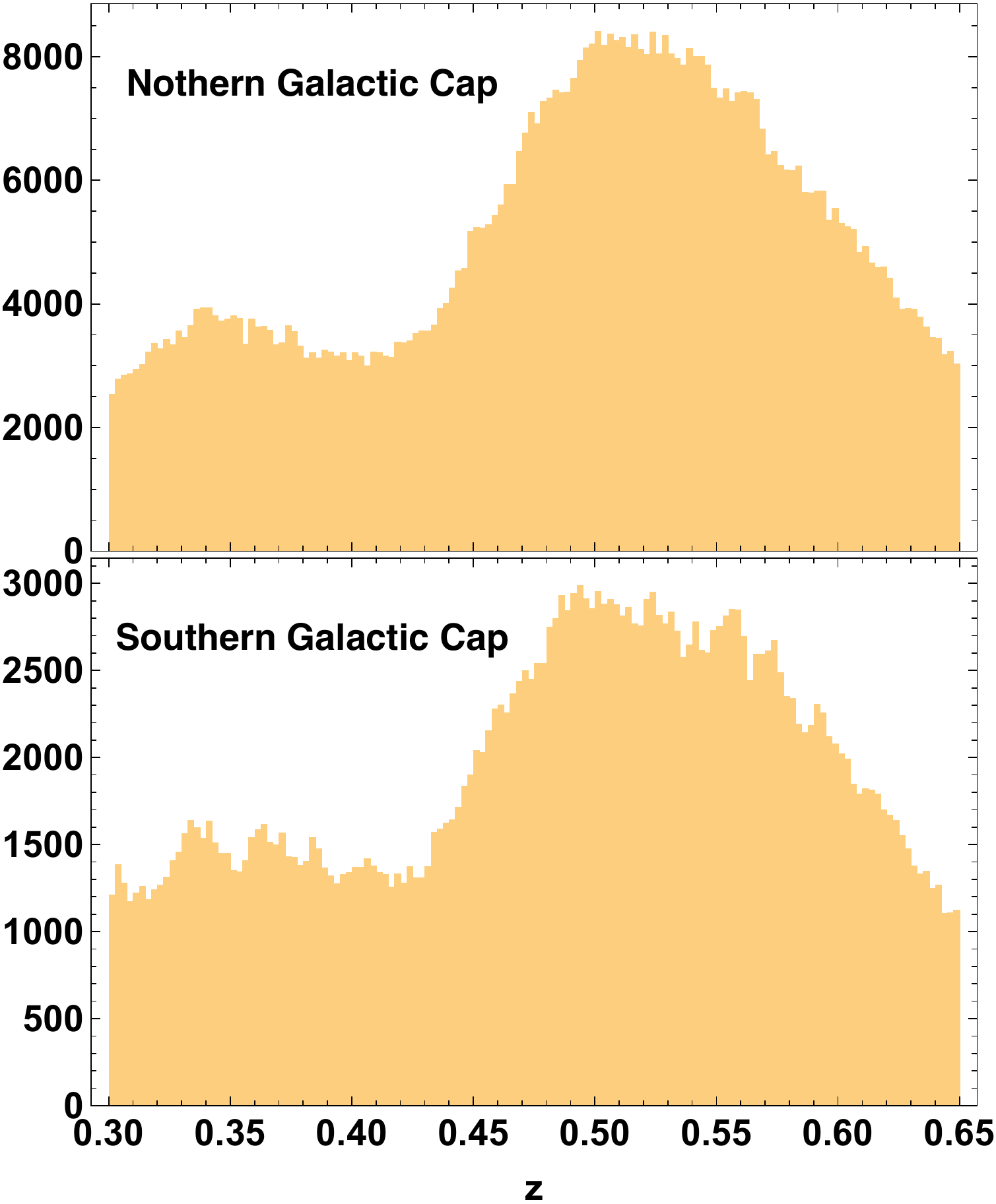}
\caption{Redshift distribution of the 720,113 galaxies of the Northern Galactic Cap (top) and the 273,117 galaxies of the Southern Galactic Cap (bottom) of the \texttt{CMASSLOWZTOT} catalog of the Data Release 12 of the BOSS survey of SDSS-III that have been used for this analysis ($0.3\le z \le 0.65$). See Table~\ref{zbins} for numerical details.
}
\label{fig:histz}
\end{figure}

The SDSS observed more than one quarter of the sky using the 2.5-m Sloan Telescope in Apache Point, New Mexico.
We use the Data Release 12 (DR12) of the BOSS survey, which is the final data release and contains all SDSS observations through July 2014 \citep{Alam:2015mbd,Reid:2015gra}. 

DR12 features the LOWZ sample of luminous red galaxies ($z\lesssim 0.4$) and the CMASS sample of massive galaxies ($0.4\lesssim z\lesssim 0.7$), in both the Northern and Southern Galactic Caps (NGC and SGC), with a total footprint of about 10,400~deg$^2$.
Figure~\ref{fig:histz} shows the redshift distributions of the 720,113 NGC galaxies and of the 273,117 SGC galaxies in the redshift interval $0.3\le z \le 0.65$.
Regarding the random catalogs, we use the \texttt{random0} versions; the NGC and the SGC have 37,115,850 and 13,647,368 random points ($0.3\le z \le 0.65$), respectively, and are 
produced in a manner independent of cosmology. The random catalogs are about 50 times larger than the corresponding real ones.
This means that the shot noise introduced by the random catalogs will be negligible as compared to the one due to the galaxy catalogs.
In order to maximize the statistical significance of the BAO signal we will consider only one bin with $0.3\le z \le 0.65$, see Table~\ref{zbins} for details. The width of the bin is similar to ones adopted in \cite{Sanchez:2012eh}.



\begin{table*}
\begin{center}
\begin{tabular}{|l|c|c|c|c|c|c|p{2.15cm}|c|}
\hline
\hline
Redshift bin & $z_{\rm eff}$ & $[z_{\rm min},z_{\rm max}]$& $\delta z$ & $\delta \phi$ & $\delta \theta$   & galaxies&   pixels with 50 or more galaxies & $\overline{\rm gal/pixel}$  \\
\hline
\hline
NGC &0.507 & $[0.30, 0.65]$ & 0.007 &$1.25^{\circ}$&$1.25^{\circ}$ & 720,113& 5829 & 124 \\
SGC & &  &&&&  273,117&1875& 146 \\
Total & &&&&  & 993,230 & 7704&129\\ 
\hline
\hline
\end{tabular}
\caption{Properties of the redshift bin adopted in the analysis. The effective redshift $z_{\rm eff}$ is defined in equation \eqref{zeff}. }
\label{zbins}
\end{center}
\end{table*}

In general, the effective redshift is defined as the weighted mean redshift of the galaxies:
\begin{equation} \label{zeff}
z_{\rm eff} = \frac{\sum_{i} d_{i} z_{i}}{\sum_{i} d_{i}} \,,
\end{equation}
where $i$ labels the $i$-th galaxy with weight $d_{i}$.
The equation above gives 0.494.
However, as we are studying  BAO, we will adopt the following more specialized estimator \citep{Beutler:2011hx}:
\begin{equation} \label{zeff2}
z_{\rm eff} = \frac{\sum_{i<j} d_{i} d_j (z_{i} + z_j)/2}{\sum_{i<j} d_{i}d_j} \simeq 0.507 \,,
\end{equation}
where the sum is over the galaxy pairs $\{i,j\}$ separated by $\Delta z_{\rm BAO}\pm \delta z$, where $\delta z$ is the redshift sub-bin discussed in the next Section.
The advantage of this definition is that it considers the weighted average redshift of the galaxy pairs that contribute to the BAO signal and so it is expected to better characterize its effective redshift, especially for the large redshift bin that we are considering.
As $\Delta z_{\rm BAO}$ it is not known in advance, we compute the $z_{\rm eff}$ of \eqref{zeff2} after having estimated $\Delta z_{\rm BAO}$.

\section{Method}\label{method}

We now extend the methodology introduced in \cite{Sanchez:2012eh} in order to apply it to observational data.
Furthermore, we introduce a new method to compute the covariance matrix directly from the data, and compare it successfully to the covariance matrix obtained from the BOSS mocks.

\subsection{Angular pixels}

Following \cite{Sanchez:2012eh}, we divide the redshift bin into  angular pixels which should be small enough in order to consider galaxies as collinear as possible and large enough in order to retain enough galaxies so that the computation of the correlation function will not be shot-noise dominated.
A large angular pixel induces a change in the slope of the correlation function at small scales, a smoothing effect produced by the inclusion of galaxy pairs that are not exactly collinear.
However, the scale at which this effect appears (fixed by the pixel size)  is very far from the BAO scale, that therefore remains unaffected.

\cite{Sanchez:2012eh} verified that one can safely use angular pixels up to 1 deg$^{2}$.
We will adopt a slightly larger angular pixel of $(1.25 \, \text{deg})^2$ as our catalog is not as dense as the $N$-body catalog used by \cite{Sanchez:2012eh}.

The correlation function is then calculated within these long and thin squared angular bins.%
\footnote{These bins look very much like the ``spaghetti alla \href{https://it.wikipedia.org/wiki/Chitarra_(gastronomia)}{chitarra}.''}
In other words, we are computing a 1-d (radial) correlation function.

\subsection{Counting galaxies with weights}\label{counts}

We will now discuss how to obtain the correlation function within the angular bin $\alpha$. 
Let us consider then the case of $n_d$ (data) galaxies $D_{i}$ with weights $d_i$ and $n_r$ (random) galaxies $R_{j}$ with weights $r_j$.
Our aim is to find at which redshift separation $\Delta z$ will we find the BAO feature. Consequently, we need to compute the correlation function using redshift bins in the redshift-separation space. We will use $N$ redshift sub-bins of $\delta z=0.007$, labelled according to $\Delta z_{\beta}$. This value was chosen in order to have enough resolution to see the BAO bump and, at the same time, have enough galaxies in each bin in order not to increase shot noise.
As we will use the optimal estimator by \cite{1993ApJ...412...64L}, we have to compute the galaxy-galaxy ($DD$), random-random ($RR$) and galaxy-random ($DR$) counts.

\subsection{Galaxy weights}\label{weights}

\begin{figure}
\centering
\includegraphics[width= \columnwidth]{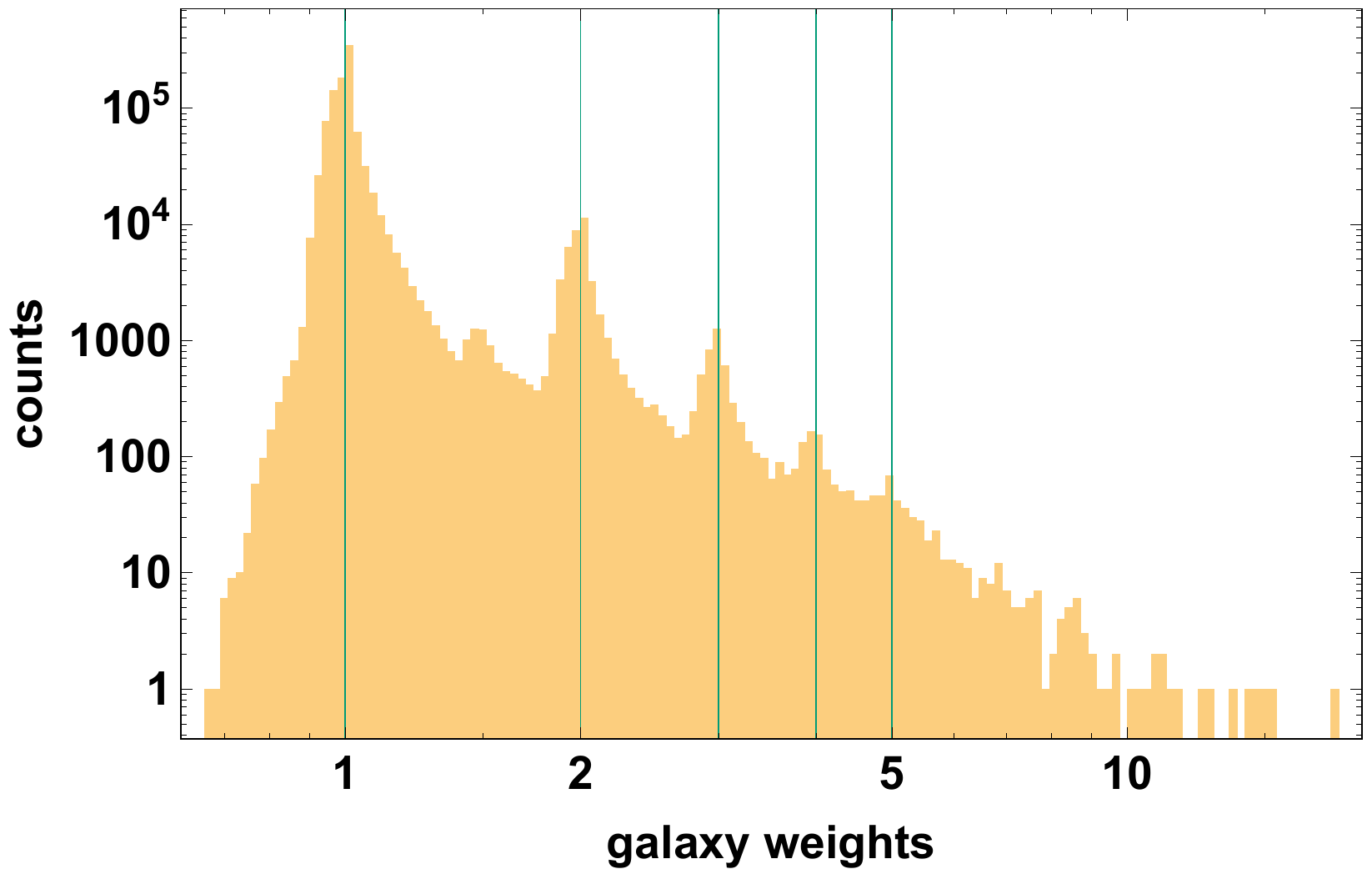}
\caption{Distribution of the galaxy weights.
}
\label{fig:g-weights}
\end{figure}

As explained in \cite{Reid:2015gra}, galaxies are weighted according to:
\begin{equation}
w_{i}= w_{{\rm sys},i} (w_{{\rm noz},i}+w_{{\rm cp},i}-1)
\end{equation}
where $w_{\rm sys}=w_{\rm star} \, w_{\rm seeing}$ is the total systematic weight, $w_{\rm noz}$ is the redshift failure weight and $w_{\rm cp}$ is the close pair weight. 
Usually, one then multiplies the above weights by the FKP weights as they are meant to optimally weight the survey galaxies \citep{Feldman:1993ky}. 
However, as the FKP weights do assume a cosmology and optimize the computation of the 3-d correlation function (we are computing a 1-d correlation function), we will carry out the analysis without these weights. Consequently, we will use:
\begin{equation}
d_{i} = w_{i}  \qquad \text{and} \qquad r_{j}=1 \,.
\end{equation}
Figure~\ref{fig:g-weights} shows the distribution of galaxy weights and is a superposition of two distributions.
The LOWZ catalog features a discrete distribution of weights at the values $\{1,2,3,4,5\}$, marked in Figure~\ref{fig:g-weights} with green gridlines. Integer values are usually due to fiber-collision upweightings.
The CMASS catalog has instead a smooth distribution, peaked $\{1, 1.5, 2,3,4,5\}$. Besides fiber-collision weights, also weights specific to this higher-$z$ catalog were used, see \cite{Reid:2015gra,Ross:2016gvb} for details.

\subsection{$DD$ or $RR$ counts}\label{DD}

The $DD$ counting proceeds as follows; the $RR$ case is analogous. One sits on the $i$-th galaxy and counts the $n_d-1$ galaxies that fall into the various redshift sub-bins. The counts are given by the $N$-dimensional vector $\tilde v_{i}$:%
%
\begin{equation}
\tilde v_{i}=d_{i} \{ \dots, \sum_{k \in \beta} d_{k}, \dots \}_{i} \,,
\end{equation}
where the $k$ data galaxies are the ones that belong to that $\beta$ sub-bin.
Clearly, the sum of the components of this vector is:
\begin{equation}
\tilde V_{i} \equiv \sum_{\beta=1}^{N}  \tilde v_{i,\beta} =d_{i} \sum_{k\neq i} d_{k} \,. \label{sct}
\end{equation}
In order to estimate $DD$ one has to stack the contributions from all the pairs; that is, one has to sum the vectors $\tilde v_{i}$ and then normalize the result:
\begin{align}
DD_{\alpha} &= \frac{\sum_{i} \tilde v_{i}}{\sum_{l}\tilde V_{l} }
=\frac{\sum_{i} \tilde v_{i}}{\sum_{l} d_{l}\sum_{k \neq l}d_{k}}   \nonumber \\
&= \frac{\sum_{i} \tilde v_{i}}{(\sum_{l} d_{l})^{2}-\sum_{l} d_{l}^{2}} \,, \label{DD}
\end{align}
where $\alpha$ marks the particular angular bin we are considering.
Note that:
\begin{equation}
\sum_{\beta=1}^{N}  DD_{\alpha \beta}=1 \,.
\end{equation}
In the case of unitary weights one has:
\begin{equation}
DD_{\alpha} =\frac{1}{n_{d}(n_{d}-1)} \sum_{i=1}^{n_{d}} \tilde v_{i} \,,
\end{equation}
where $\tilde v_{i}$ becomes a vector of integer counts.

If one, more efficiently, counts pairs only once (i.e.~not both $d_{1}d_{3}$ and $d_{3}d_{1}$), then the counts are given by the vector~$\hat v_{i}$:%
%
\begin{equation}
\hat v_{i}=d_{i} \{ \dots, \sum_{k \in \beta, k>i} d_{k}, \dots \}_{i} \,.
\end{equation}
Consequently:
\begin{align}
DD_{\alpha} &= \frac{\sum_{i} \hat v_{i}}{\sum_{l}\hat V_{l} }
=\frac{\sum_{i} \hat v_{i}}{\sum_{l} d_{l}\sum_{k > l}d_{k}}   \nonumber \\
&= \frac{\sum_{l} \hat v_{l}}{[(\sum_{l} d_{l})^{2}-\sum_{l} d_{l}^{2}]/2} \,, \label{eqdd2}
\end{align}
which in the case of unitary weights becomes:
\begin{equation}
DD_{\alpha} =\frac{1}{n_{d}(n_{d}-1)/2} \sum_{i=1}^{n_{d}} \hat v_{i} \,.
\end{equation}
The latter equation features the standard normalization. Equation~\eqref{eqdd2} gives the same counts as compared with equation~\eqref{DD}.

For later use we define the total weight of the angular pixel $\alpha$ as:
\begin{equation} \label{nuga1}
dd_{\alpha}=  \frac{(\sum_{l} d_{l})^{2}-\sum_{l} d_{l}^{2}}{2} \,,
\end{equation}
which generalizes the total number of different pairs in the angular pixel $\alpha$:
\begin{equation} \label{nuga2}
n_{dd,\alpha}= \frac{n_{d}(n_{d}-1)}{2} \,.
\end{equation}
%

\subsection{$DR$ counts}\label{DR}

The $DR$ counting proceeds as follows; the $RD$ case gives exactly the same result.%
\footnote{The $RD$ computation is numerically slower because there are more random galaxies than real ones so that one has to sit on more galaxies. The counting is performed vectorially. This behavior may depend on the way the calculation was implemented.}
The index $i$ will label the data galaxies while the index $j$ the random galaxies.
One sits on the $i$-th galaxy and counts the $n_r$ galaxies that fall into the various redshift sub-bins.
The counts are given by the $N$-dimensional vector $\tilde u_{i}$:%
%
\begin{equation}
\tilde u_{i}=d_{i} \{ \dots, \sum_{k \in \beta} r_{k}, \dots \}_{i} \,,
\end{equation}
where the $k$ random galaxies are the ones that belong to that $\beta$ sub-bin.
The sum of the components of $\tilde u_{i}$ is:
\begin{equation}
\tilde U_{i} \equiv \sum_{\beta=1}^{N} \tilde u_{i,\beta} = d_{i} \sum_{k} r_{k} \equiv d_{i} \, r_{\rm tot}  \,,
\end{equation}
where we have defined the total weight $r_{\rm tot}$.
In order to estimate $DR$, one has to stack the contributions from all the pairs; that is, one has to sum the vectors $\tilde u_{i}$ and then normalize the result:
\begin{align}
DR_{\alpha} &= \frac{\sum_{i} \tilde u_{i}}{\sum_{l}\tilde U_{l} }
=\frac{\sum_{i} \tilde u_{i}}{\sum_{l} d_{l}\, r_{\rm tot} } 
=\frac{\sum_{i} \tilde u_{i}}{d_{\rm tot} \, r_{\rm tot} }  \,. \label{DR}
\end{align}
Note that:
\begin{equation}
\sum_{\beta=1}^{N}  DR_{\alpha \beta}=1 \,. \label{norma2}
\end{equation}
In the case of unitary weights one has:
\begin{equation}
DR_{\alpha} =\frac{1}{n_{d}\, n_{r}} \sum_{i} \tilde u_{i} \,, \label{DR0}
\end{equation}
where $\tilde u_{i}$ becomes a vector of integer counts. The latter equation features the standard normalization.

For later use we define the total weight of the angular pixel $\alpha$ as:
\begin{equation}
dr_{\alpha}=  \sum_{i} d_{i} \sum_{j} r_{j} = d_{\rm tot} \, r_{\rm tot} \,,
\end{equation}
which generalizes the total number of different pairs in the angular pixel $\alpha$:
\begin{equation}
n_{dr,\alpha}= n_{d} \, n_{r} \,.
\end{equation}
%

\subsection{Correlation function and covariance matrix}\label{pixel}

After carrying out the procedure above we obtain the following quantities:
\begin{align}
& DD_{\alpha \beta}   \phantom{bacani}  dd_{\alpha}   \phantom{bacano}  n_{dd,\alpha} \,, \nonumber \\
& RR_{\alpha \beta}   \phantom{bacana}  rr_{\alpha}     \phantom{bacana}  n_{rr,\alpha}  \,, \nonumber \\
& DR_{\alpha \beta}   \phantom{bacano}  dr_{\alpha}    \phantom{bacano}  n_{dr,\alpha} \,,
\end{align}
where $\alpha$ labels an angular pixes in the Northern or Southern Galactic Cap and $\beta$ labels the redshift sub-bins.

\begin{figure}
\centering 
\includegraphics[width= \columnwidth]{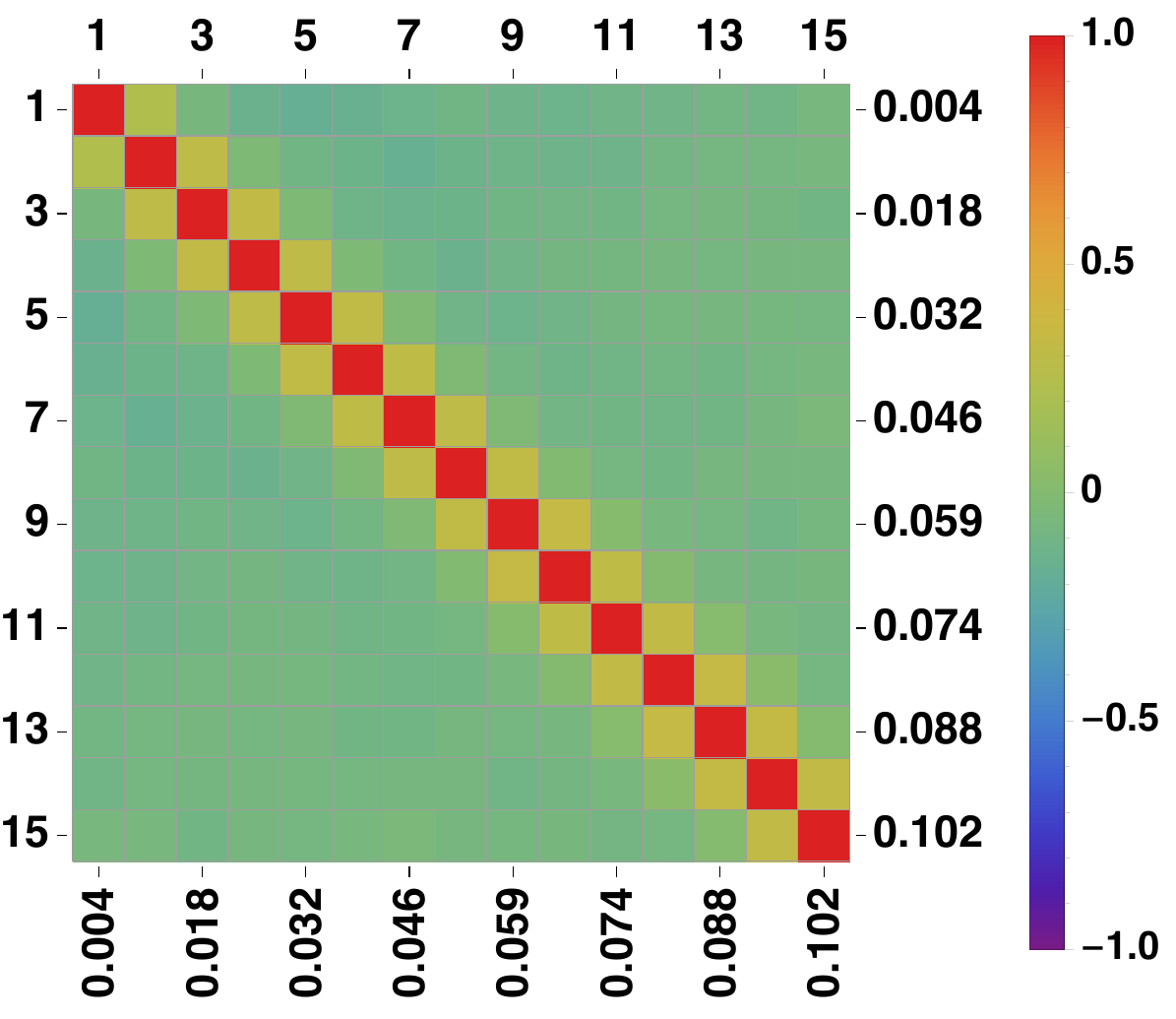}
\caption{Correlation matrix relative to the covariance matrix $\Sigma_{\beta \beta'}$ of equation \eqref{covame}. See Table~\ref{zbins} for the bin specifications.}
\label{fig:cova}
\end{figure}

We compute the correlation function directly in every pixel $\alpha$ and then calculate the mean vector and its $N\times N$ covariance matrix:
\begin{align}
\xi_{\alpha \beta} &\equiv \frac{DD_{\alpha \beta}-2DR_{\alpha \beta}+RR_{\alpha \beta}}{RR_{\alpha \beta}} \,, \\
\xi_{\beta} &= \left \langle \xi_{\alpha \beta} \right \rangle_{\alpha} \,, \label{xime} \\
\Sigma_{\beta \beta'} &= \frac{\text{cov} (\xi_{\alpha \beta})}{n_{\rm eff}} \,. \label{covame}
\end{align}
where we used for $\xi_{\alpha \beta}$ the optimal estimator by \cite{1993ApJ...412...64L}.
As we are not mixing pixels before the computation of the correlation function, we merge Northern and Southern Galactic Cap pixels.
The correlation matrix is shown in Figure~\ref{fig:cova}, while the mean correlation function and the diagonal of the covariance matrix are shown in Figure~\ref{fig:xi}.

\begin{figure}
\centering
\includegraphics[width= \columnwidth]{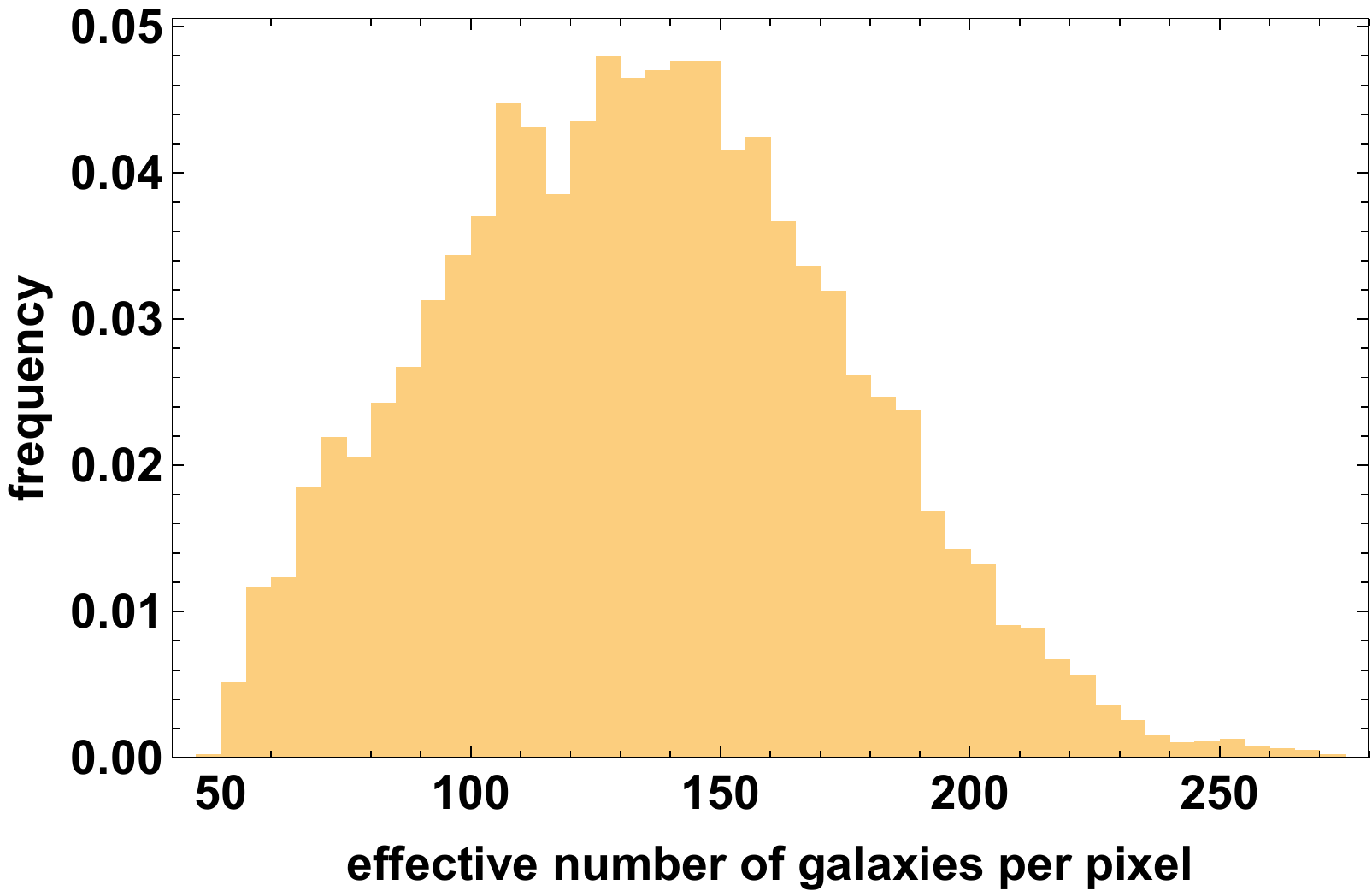}
\caption{Distribution of the effective number of galaxies per pixel $n^{\rm eff}_{dd,\alpha}\simeq \sqrt{2 dd_{\alpha}}$, see equations (\ref{nuga1}-\ref{nuga2})}
\label{fig:weights}
\end{figure}

As the angular pixels feature a wide range of number of galaxies, they do not all have the same statistical constraining power: see Figure~\ref{fig:weights} where  the effective number of galaxies per pixel $n^{\rm eff}_{dd,\alpha}\simeq \sqrt{2 dd_{\alpha}}$ is shown.
Therefore, in order to obtain smaller errors and higher significance,  it seems appropriate to weight the computation of the average $\xi_{\beta}$ and the covariance $\Sigma_{\beta \beta'}$ according to the number of galaxies present in the angular pixel.
As angular pixel weights we adopted the weights $rr_{\alpha}$.
The reason is that the random catalog closely follows the data catalog by construction and, at the same time, the weights $rr_{\alpha}$ are uncorrelated with the measurements so that this weighing scheme should not cause bias or increased variance.
We leave to future work the study of an optimal weighting scheme akin to the one proposed by \cite{Feldman:1993ky} in order to balance sample variance and shot noise.

The covariance matrix of equation \eqref{covame} is relative to the mean vector $\xi_{\beta}$. That is why it is divided by the effective number of pixels $n_{\rm eff}$:
\begin{equation}
n_{\rm eff} = \frac{\left(\sum_{\alpha} rr_{\alpha}\right)^{2}}{\sum_{\alpha} rr_{\alpha}^{2}} \simeq 6671 \,.
\end{equation}
%

\subsection{MultiDark-Patchy mocks covariance matrix}\label{mocks}

The determination of the covariance matrix \eqref{covame} directly from the data is carried out under the approximation that $\xi_{\alpha \beta}$ and $\xi_{\alpha' \beta}$ are independent for $\alpha \neq \alpha'$.
However, possible correlations could affect this estimation, possibly producing smaller errors and so artificially boosting the significance.
In this specific case, since the BAO scale corresponds to $\sim 5$ deg 
at $z\sim 0.5$ \citep[see, e.g.,][figure 3]{vonMarttens:2018bvz}, 1.25 deg pixels are expected to be correlated by sample variance.

In order to check this important issue we computed the correlation $\xi_{\beta}$ of equation \eqref{xime} for the 500 MultiDark-Patchy mocks made for the BOSS data.
Specifically, we used the first 500 mocks of the sets \texttt{Patchy-Mocks-DR12NGC-COMPSAM\_V6C} and \texttt{Patchy-Mocks-DR12SGC-COMPSAM\_V6C}, and also the 50 times as big corresponding random catalog.
See \citet{Kitaura:2015uqa,Rodriguez-Torres:2015vqa} for details regarding these mocks.

\begin{figure}
\centering 
\includegraphics[width= \columnwidth]{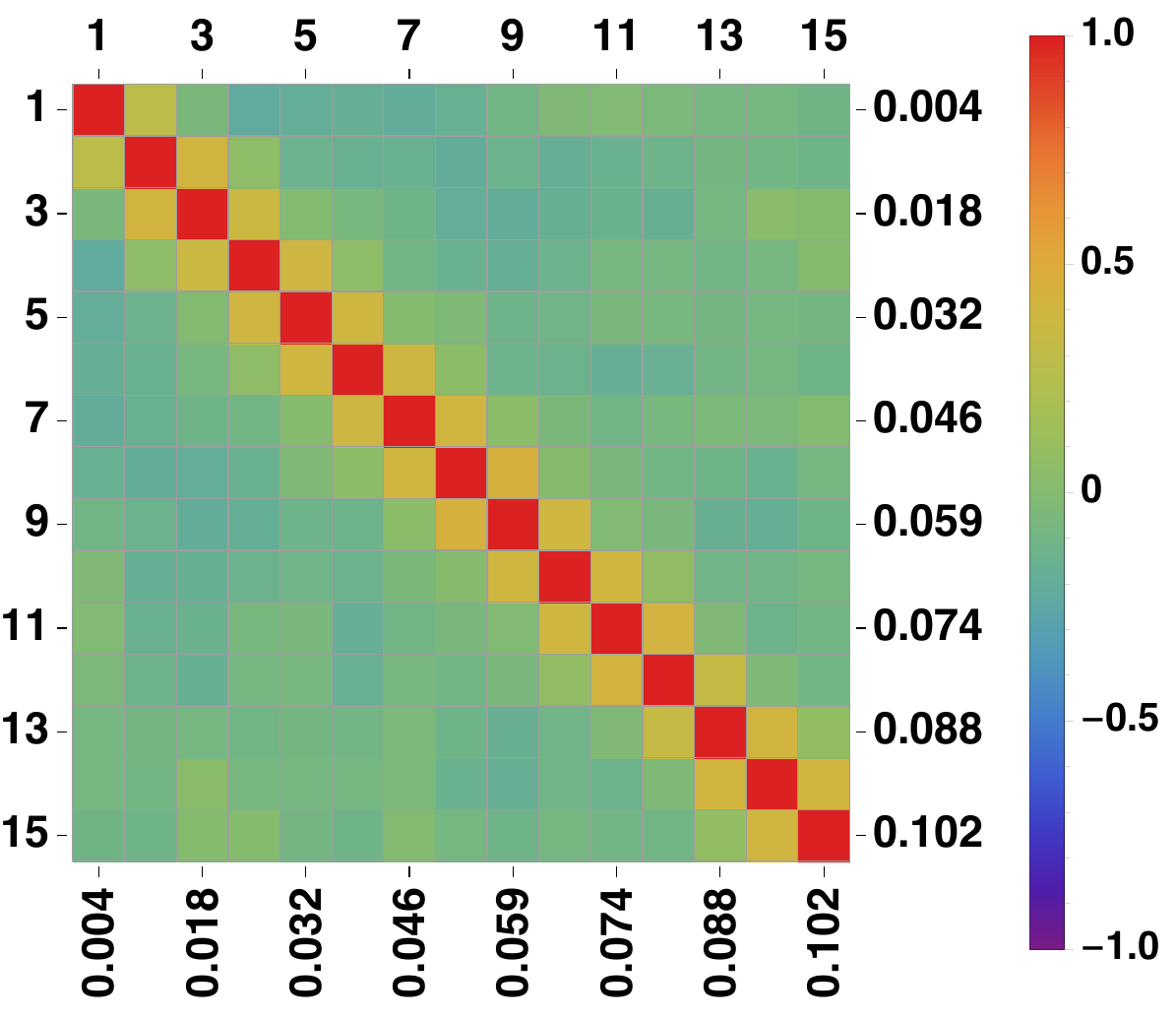}
\includegraphics[width= \columnwidth]{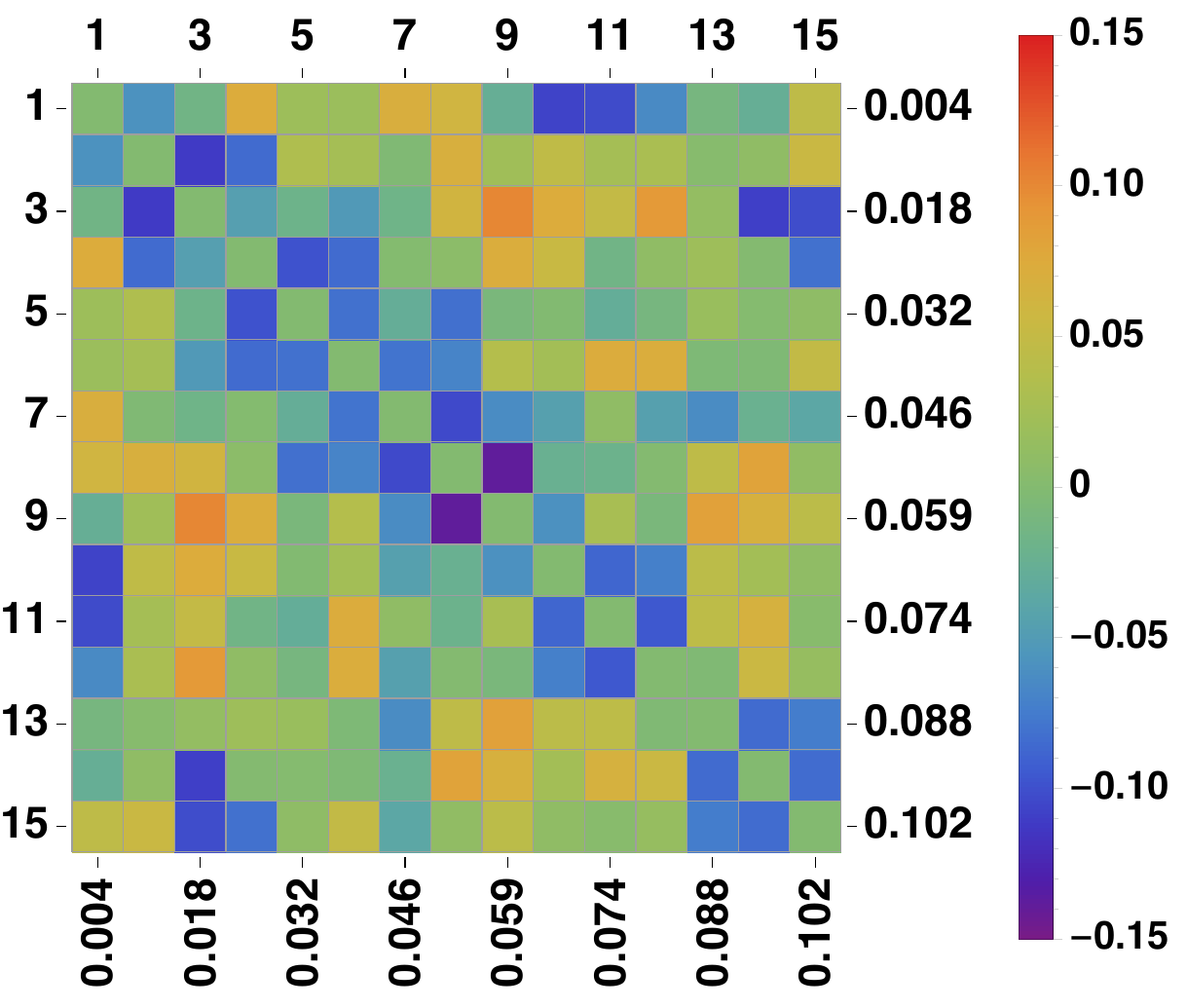}
\caption{Correlation matrix relative to the MultiDark-Patchy mocks (top) and difference with respect to the correlation matrix of Figure~\ref{fig:cova} which was estimated directly from the data (bottom).}
\label{fig:covamo}
\end{figure}

\begin{figure}
\centering 
\includegraphics[width= \columnwidth]{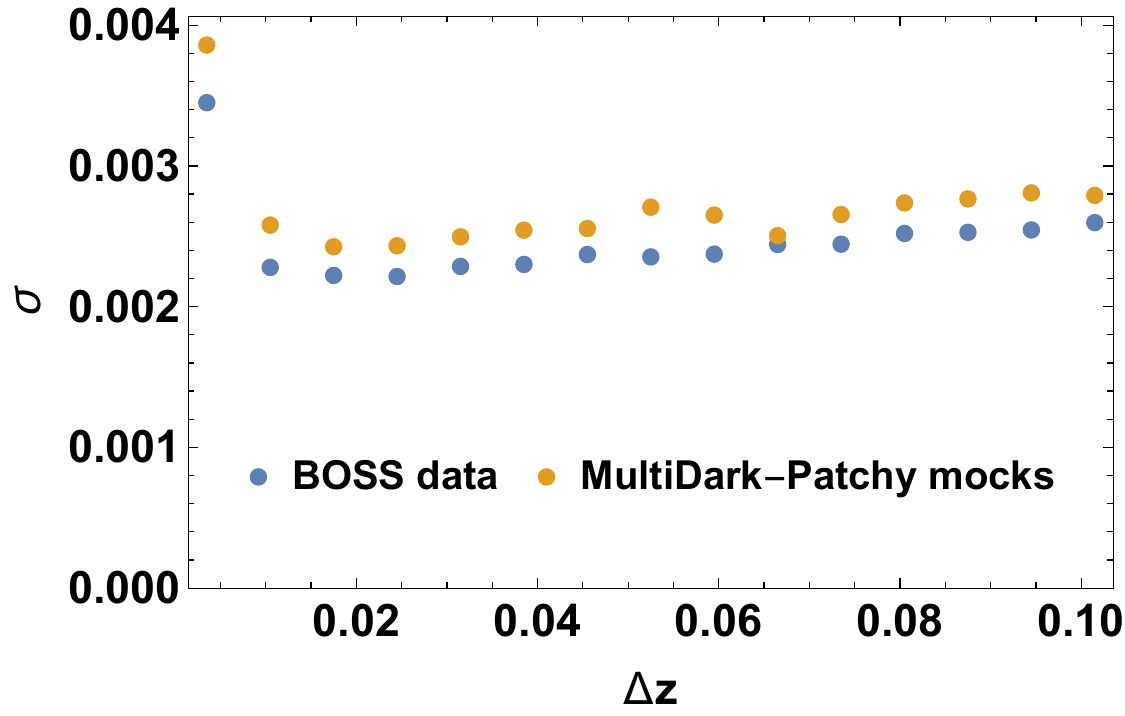}
\caption{Square root of the diagonal of the covariance matrix estimated from the data and of the diagonal of the covariance matrix from the MultiDark-Patchy mocks.}
\label{fig:covadia}
\end{figure}

We then computed the mock covariance matrix $\Sigma^{\rm mocks}_{\beta \beta'}$, which is shown via the correlation matrix in Figure~\ref{fig:covamo} and via its diagonal in Figure~\ref{fig:covadia}.
The bottom panel shows the difference between the data correlation matrix and the one obtained from the BOSS mocks. The overall agreement is good, even if one may notice that the correlation between nearby redshift sub-bins is a little underestimated when computed from the data.
In Figure~\ref{fig:covadia} we compare the errors: the mock covariance matrix features errors only 9\% larger, a very good agreement.

We conclude that, for the special case of the fiducial $\Lambda$CDM cosmology, the covariance matrix estimated from the data with the novel procedure of Section~\ref{pixel} is trustworthy.
Therefore, in the main analysis we will use the covariance matrix \eqref{covame} estimated directly from the data in order to keep the analysis model independent. We will adopt the mock covariance matrix in Section~\ref{mockcovare}.

\subsection{BAO position}\label{bao}

Within the homogeneous and isotropic FLRW model, the BAO bump in redshift space is given by~\citep{Hogg:1999ad}:
\begin{equation}
\Delta z_{\rm BAO}^{\rm theo} = \frac{r_{d} \, H(z)}{c} \,,
\end{equation}
where $r_{d}$ is the line-of-sight comoving sound horizon at the drag epoch, $H$ is the Hubble rate and $c$ is the speed of light.
Using the fiducial cosmology of \citealt{Alam:2015mbd} ($\Omega_{m}=0.31$, $h=0.676$ and $\Omega_{b}h^{2}=0.022$) one  obtains the theoretical prediction ($r_{d}=147.78$ Mpc):
\begin{equation} \label{Dzfid}
\Delta z^{\rm fid}_{\rm BAO}(z=0.507)\simeq0.04410 \,.
\end{equation}

However, as we are considering a large redshift bin ($0.3\le z \le 0.65$), one may obtain a more accurate prediction by adopting a method similar to the one used in equation~\eqref{zeff2}:
\begin{equation} \label{Dzfid2}
\Delta z^{\rm fid}_{\rm BAO} = \frac{\sum_{i<j} d_{i} d_j \, \Delta z_{\rm BAO}^{\rm theo}(\frac{z_i + z_j}{2})}{\sum_{i<j} d_{i}d_j} \simeq 0.04414 \,,
\end{equation}
where the sum is over the galaxy pairs $\{i,j\}$ separated by $\Delta z_{\rm BAO}\pm \delta z$.
As $\Delta z_{\rm BAO}$ it is not known in advance, we compute the $\Delta z^{\rm fid}_{\rm BAO}$ of \eqref{Dzfid2} after having estimated $\Delta z_{\rm BAO}$.

As we expect to detect the BAO feature at $\Delta z \approx 0.05$, we will restrict the analysis to a range of $0.01 \lesssim \Delta z \lesssim 0.1$ in order to reduce the impact of poorly sampled regions of the correlation function.

\section{Results} \label{results}

\begin{figure}
\centering
\includegraphics[width=\columnwidth]{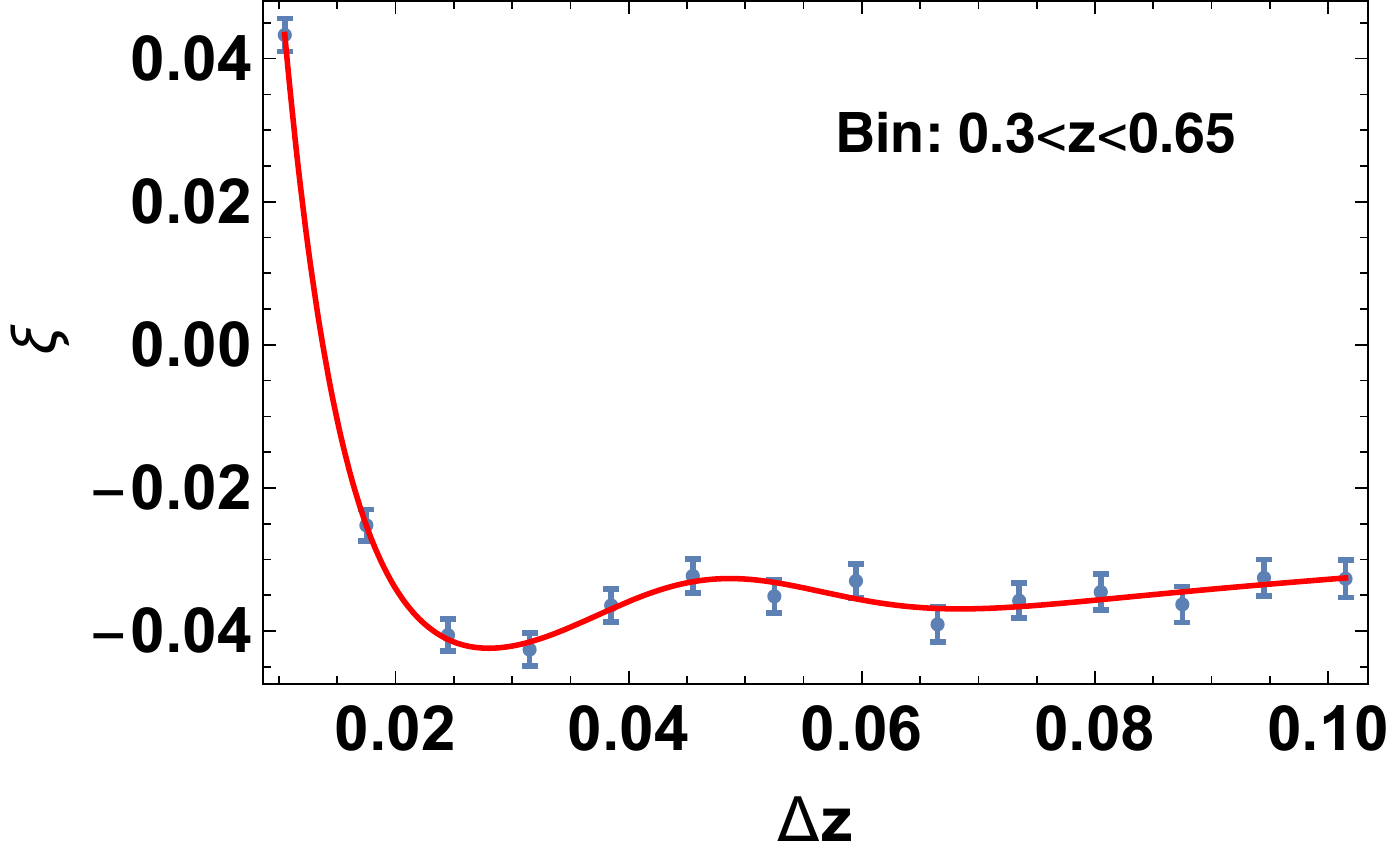}
\caption{Correlation function $\xi_{\beta}$ together with the fit to equation \eqref{para}.
The error bars are from the diagonal elements of the covariance matrices; however, the points are correlated as visible in Figure~\ref{fig:cova}.
}
\label{fig:xi}
\end{figure}

In order to determine $\Delta z_{\rm BAO}$ we will use the phenomenological parametrization proposed by \cite{Sanchez:2012eh}:
\begin{align} \label{para}
\xi_{\parallel}(\Delta z) \!=\! A+  B e^{-C \Delta z} \!- D e^{-E \Delta z} \!+ F e^{-\frac{(\Delta z - \Delta z_{\rm BAO})^{2}}{2\sigma^2}},
\end{align}
which was proven to be flexible enough so as to produce an unbiased estimation of $\Delta z_{\rm BAO}$.

We define the following $\chi^{2}$ function:
\begin{equation}
\chi^{2} = \{\xi_{\parallel}(\Delta z_{\beta})-\xi_{\beta}\}\Sigma^{-1}_{\beta \beta'} \{\xi_{\parallel}(\Delta z_{\beta'})-\xi_{\beta'}\} \,, \label{chi21}
\end{equation}
where repeated indexes are summed over and $\xi_{\beta}$ is the one of equation~\eqref{xime} and shown in Figure~\ref{fig:xi}.
The best-fit model is shown with a red line in Figure~\ref{fig:xi}.
The $\Delta z_{\rm BAO}$ parameter describes the position of the BAO feature in redshift space, while the cosmological interpretation of the other parameters is limited, since equation \eqref{para} is an empirical description valid only in the neighborhood of the BAO peak.

\begin{figure*}
\centering
\includegraphics[width=\textwidth]{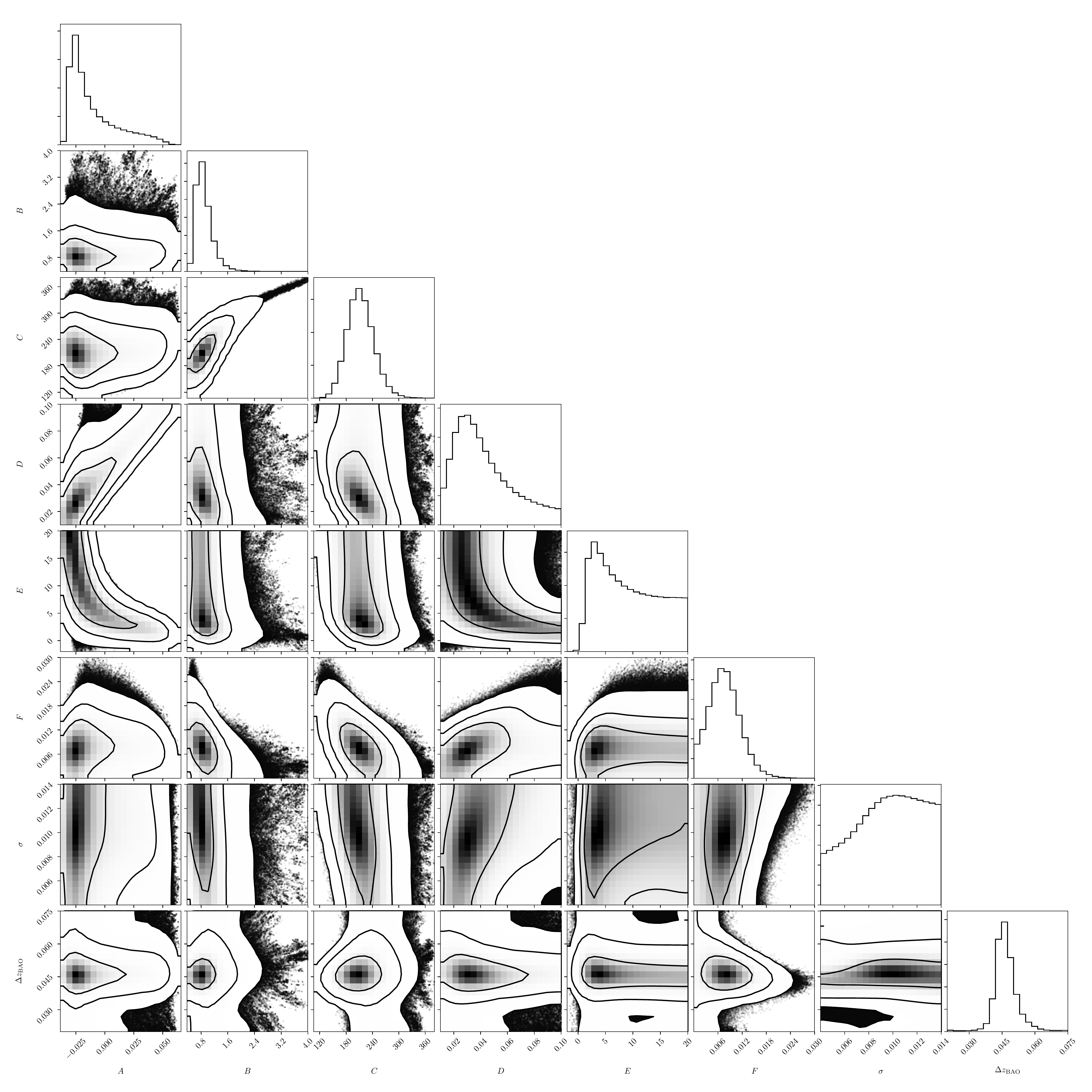}
\caption{Triangular plot relative to the $\chi^{2}$ function of equation~\ref{chi21}. The contours contain 1-, 2- and 3-$\sigma$ of the total probability.
}
\label{fig:corner}
\end{figure*}

In order to determine the statistical error on $\Delta z_{\rm BAO}$ we perform an MCMC analysis using \texttt{emcee} \citep{ForemanMackey:2012ig} in the 8-dimensional parameter space spanned by $\{ A, B, C, D, E, F, \sigma, \Delta z_{\rm BAO} \}$.
We adopted flat priors on all parameter and we constrained $F>0$ as the BAO signal is expected to be a peak.
Also, we restrict the width of the peak to $0.004<\sigma<0.014$, because a tight peak cannot be constrained due to the limited resolution provided by the redshift sub-bins and a broad peak would exceed the width of the data.
Also, we restrict the position of the peak to be within the data rage, $0.02<\Delta z_{\rm BAO}<0.075$.
The result of this analysis is shown in Figure~\ref{fig:corner}. Interestingly, we can observe little correlation between $\Delta z_{\rm BAO}$ and the other parameters.

The previous analysis gives the following model-independent BAO estimation:
\begin{align}
\Delta z_{\rm BAO}(z_{\rm eff}\!=\!0.507)=
0.0456_{-0.0038}^{+0.0045} \simeq 0.0456 \pm 0.0042  , \label{mibao}
\end{align}
which agrees with the value relative to the fiducial cosmology of \cite{Alam:2015mbd} given in equation \eqref{Dzfid2}.
\cite{Sanchez:2012eh} estimated a total systematic error of 0.33\%, which is much smaller than the 9\% statistical error in equation \eqref{mibao} and can be neglected.

In order to assess the significance of the detection of the BAO peak, we compare the $\chi^{2}_{\rm min}$ relative to equation~\eqref{chi21} (6 d.o.f.) with the $\chi^{2}_{\rm min}$ relative to the ``noBAO'' model, that is, equation~\eqref{chi21} with $F=0$  (9 d.o.f.). Comparing these two values provides a measure of our level of confidence that the BAO feature exists in the data. This analysis gives:
\begin{equation}
\chi^{2}_{\rm min}(\text{noBAO})-\chi^{2}_{\rm min}(\text{BAO}) = 12.4- 7.9=4.5 \,.
\end{equation}
As one expects $\chi^{2}_{\rm min}(\text{noBAO})$ to be 3 points higher owing to the increased d.o.f., the significance of the detection is small. Alternatively, the noBAO model is rejected with a $p$-value of only 0.2.
Statistical errors are significantly higher and significance is lower as compared to the analysis that uses the 3-d correlation function
because cross-pixel correlations cannot be used: this is the price of dropping completely basic assumptions in order to carry out a fully model-independent analysis.
Future catalogs are expected to yield higher significance results.

\subsection{Mock data analysis} \label{mockcovare}

Here, in order to test the robustness of the method we repeat the analysis using the mock covariance matrix estimated in Section~\ref{mocks}. The result is:
\begin{align}
&\Delta z_{\rm BAO}= 0.0442 \pm 0.0044  \,,\\
&\chi^{2}_{\rm min}(\text{noBAO})-\chi^{2}_{\rm min}(\text{BAO}) = 13.7- 9.4=4.3 \,,
\end{align}
in good agreement with the main result that uses the covariance matrix obtained directly from the data.

Finally, we performed the analysis using the mean correlation function from the mocks and the mock covariance matrix relative to the mean, defined as $\Sigma^{\rm mocks}_{\beta \beta'}/500$.
The result is:
\begin{align}
&\Delta z_{\rm BAO}=0.04437 \pm 0.00024  \,,\\
&\chi^{2}_{\rm min}(\text{noBAO})-\chi^{2}_{\rm min}(\text{BAO}) = 562- 7=555 \,.\,
\end{align}
which shows that the phenomenological parametrization of equation \eqref{para} is a good model.
The position of the BAO peak is less than 1$\sigma$ from the value relative to the fiducial cosmology give in equation~\eqref{Dzfid2}.
This small bias is inconsequential for our analysis as it amounts to an offset of $0.06 \sigma$ in equation \eqref{mibao}.
This bias is possibly due to the binning strategy and weighting scheme used.

\section{Conclusions} \label{conclusions}

In this work, using almost one million galaxies from the final data release of the BOSS survey (SDSS-III DR12), we have obtained, albeit with low significance, a model-independent determination of the radial BAO feature, see equation \eqref{mibao}.
It features a 9\% error and can be used to constrain the background expansion of exotic models for which the assumptions adopted in the standard analysis of \cite{Alam:2016hwk} cannot be satisfied.
This is the case, for example, of inhomogeneous models that are used to test homogeneity without assuming it~\citep{Valkenburg:2012td}.

It is worth stressing that we have introduced a new method to compute the covariance matrix directly from the data, and validated it against the covariance matrix obtained from the BOSS mocks. Consequently, following the method presented in this paper, it is possible to obtain fully model-independent BAO constraints.

Future galaxy catalogs from J-PAS~\citep{Benitez:2014ibt}, DESI~(\href{https://www.desi.lbl.gov/final-design-report/}{desi.lbl.gov}, \citealt{Martini:2018kdj}) and Euclid~\citep{Amendola:2016saw} will allow us to obtain high-significance model-independent determinations of the BAO peak at several redshift values. Using the formalism here presented it will be straightforward to obtain the position of the BAO feature in different angular regions so that isotropy of the universe could be directly tested.
We conclude stressing that it is imperative to test the standard paradigm in a model-independent way in order to test its foundations, maximize the extraction of information from the data, and look for clues regarding the poorly understood dark energy and dark matter.

\section*{Acknowledgements}
It is a pleasure to thank Stefano Anselmi, David Camarena, Gabriela C.~Carvalho, Martin Crocce, Oliver Piattella, Ashley Ross  for useful comments and discussions.
A special thank to Álefe Almeida and Renan Oliveira for help on numerical aspects of this work.
VM thanks CNPq and FAPES for partial financial support.
ECI thanks CAPES for financial support.

Funding for SDSS-III has been provided by the Alfred P.~Sloan Foundation, the Participating Institutions, the National Science Foundation, and the U.S. Department of Energy Office of Science. The SDSS-III web site is \href{http://www.sdss3.org/}{sdss3.org}. 
SDSS-III is managed by the Astrophysical Research Consortium for the Participating Institutions of the SDSS-III Collaboration including the University of Arizona, the Brazilian Participation Group, Brookhaven National Laboratory, Carnegie Mellon University, University of Florida, the French Participation Group, the German Participation Group, Harvard University, the Instituto de Astrofisica de Canarias, the Michigan State/Notre Dame/JINA Participation Group, Johns Hopkins University, Lawrence Berkeley National Laboratory, Max Planck Institute for Astrophysics, Max Planck Institute for Extraterrestrial Physics, New Mexico State University, New York University, Ohio State University, Pennsylvania State University, University of Portsmouth, Princeton University, the Spanish Participation Group, University of Tokyo, University of Utah, Vanderbilt University, University of Virginia, University of Washington, and Yale University.

The massive production of all MultiDark-Patchy mocks for the BOSS Final Data Release has been performed at the BSC Marenostrum supercomputer, the Hydra cluster at the Instituto de Fısica Teorica UAM/CSIC, and NERSC at the Lawrence Berkeley National Laboratory. We acknowledge support from the Spanish MICINNs Consolider-Ingenio 2010 Programme under grant MultiDark CSD2009-00064, MINECO Centro de Excelencia Severo Ochoa Programme under grant SEV- 2012-0249, and grant AYA2014-60641-C2-1-P. The MultiDark-Patchy mocks was an effort led from the IFT UAM-CSIC by F.~Prada’s group (C.-H.~Chuang, S.~Rodriguez-Torres and C.~Scoccola) in collaboration with C.~Zhao (Tsinghua U.), F.-S.~Kitaura (AIP), A.~Klypin (NMSU), G.~Yepes (UAM), and the BOSS galaxy clustering working group.


\bibliography{references}


\bsp	
\label{lastpage}
\end{document}